\documentclass[%
superscriptaddress,
preprint,
reprint,
onecolumn,
amsmath,amssymb,
aip,
floatfix,
]{revtex4-1}

\usepackage{graphicx}
\usepackage{dcolumn}
\usepackage{bm}
\usepackage{hyperref}
\usepackage[acronym]{glossaries}
\usepackage[caption=false]{subfig}

\renewcommand{\vec}[1]{\boldsymbol{#1}}
\newcommand*{\diff}{\mathop{}\!\mathrm{d}}
\newcommand*\chem[1]{\ensuremath{\mathrm{#1}}}

\newcommand*\mean[1]{\overline{#1}}
\newcommand*{\me}{\mathrm{e}}

\newcommand{\objective}{\varphi}

\newacronym{rdf}{RDF}{radial distribution function}
\newacronym{ipp}{IPP}{isotropic pair potential}
\newacronym{bod}{BOD}{bond-orientational order diagram}
\newacronym{md}{MD}{molecular dynamics}
\newacronym{ibi}{IBI}{Iterative Boltzmann Inversion}
\newacronym{gcm}{GCM}{Gaussian core model}
\newacronym{ipl}{IPL}{inverse-power-law potential}
\newacronym{srs}{SRS}{soft-repulsive-shoulder potential}
\newacronym{sral}{SRAL}{short-range-attractive-long-range-repulsive}

\newacronym{remin}{REM}{relative entropy minimization}
\newacronym{ffrem}{FF-REM}{Fourier-filtered relative entropy minimization}
\newacronym{rsrem}{RS-REM}{real space smoothened relative entropy minimization}

\newacronym{cP1}{$cP1$}{simple cubic}
\newacronym{cI2}{$cI2$}{body-centered cubic}
\newacronym{cF4}{$cF4$}{face-centered cubic}
\newacronym{betaMn}{$cP20$}{$\beta$-manganese $cP20$-\chem{Mn}}
\newacronym{betaSn}{$tI4$}{$\beta$-tin $tI4$-\chem{Sn}}
\newacronym{sigmaphase}{$tP30$}{$\sigma$-phase $tP30$-\chem{CrFe}}
\newacronym{cF8}{$cF8$}{diamond}
\newacronym{cP4}{$cP4$}{$cP4$-\chem{Li}}
\newacronym{cP8}{$cP8$}{A15-type $cP8$-\chem{Cr_3Si}}
\newacronym{cI52}{$cI52$}{$\gamma$-brass $cI52$-\chem{Cu_5Zn_8}}
\newacronym{cP54}{$cP54$}{clathrate-I $cP54$-\chem{K_4Si_{23}}}

\begin{document}

\title{Inverse Design of Simple Pair Potentials for the Self-Assembly of Complex Structures}

\author{Carl S. Adorf}
\email{csadorf@umich.edu}
\affiliation{%
    Department of Chemical Engineering, University of Michigan, Ann Arbor 48109, USA}%
\author{James Antonaglia}
\email{jamesaan@umich.edu}
\affiliation{Department of Physics, University of Michigan, Ann Arbor 48109, USA}
\author{Julia Dshemuchadse}%
\email{djulia@umich.edu}
\affiliation{%
    Department of Chemical Engineering, University of Michigan, Ann Arbor 48109, USA}%
\email{djulia@umich.edu}
\author{Sharon C. Glotzer}
\email{sglotzer@umich.edu}
\affiliation{%
    Department of Chemical Engineering, University of Michigan, Ann Arbor 48109, USA}%
\affiliation{Department of Physics, University of Michigan, Ann Arbor 48109, USA}
\affiliation{Department of Materials Science and Engineering, University of Michigan, Ann Arbor 48109, USA}
\affiliation{%
    Biointerfaces Institute, University of Michigan, Ann Arbor 48109, USA}%

\date{\today}
             
\begin{abstract}
The synthesis of complex materials through the self-assembly of particles at the nanoscale provides opportunities for the realization of novel material properties.
However, the inverse design process to create experimentally feasible interparticle interaction strategies is uniquely challenging.
Standard methods for the optimization of isotropic pair potentials tend toward overfitting, resulting in solutions with too many features and length scales that are challenging to map to mechanistic models.
Here we introduce a method for the optimization of simple pair potentials that minimizes the relative entropy of the complex target structure while directly considering only those length scales most relevant for self-assembly.
Our approach maximizes the relative information of a target pair distribution function with respect to an ansatz distribution function \textit{via} an iterative update process.
During this process, we filter high frequencies from the Fourier spectrum of the pair potential, resulting in interaction potentials that are smoother and simpler in real space, and therefore likely easier to make.
We show that pair potentials obtained by this method assemble their target structure more robustly with respect to optimization method parameters than potentials optimized without filtering.
\end{abstract}

\maketitle

\section{Introduction}

\begin{figure*}
    \centering
    \includegraphics[width=0.60\linewidth]{{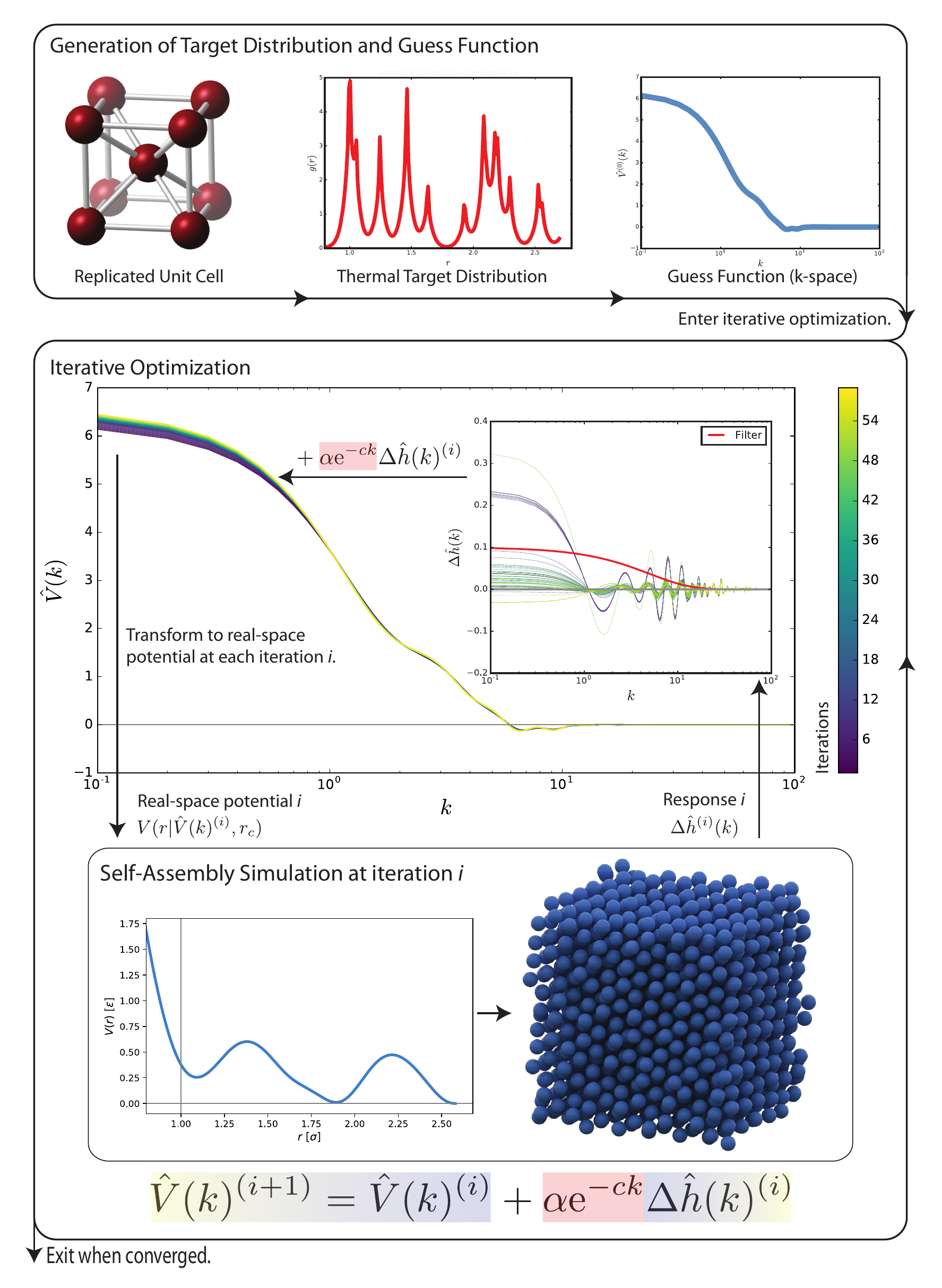}}
    \caption{
        To generate an \glsfirst{ipp} for the self-assembly of complex structures, the \glsfirst{rdf} is measured from a thermalized ideal crystal, from which we generate a smooth guess function in $k$-space.
        This guess function is then iteratively updated by transforming the potential into real space at each iteration, executing a self-assembly simulation, measuring the response, and then updating the potential accordingly in Fourier space.
        The initial guess, as well as all updates are smoothened \textit{via} a low-pass filter (shown in red) in order to ensure that the optimization is biased toward smoother potentials that carry only those length scales that are crucial for the assembly of the target structure.
    }
    \label{fig:method}
\end{figure*}

The ability to synthesize novel complex materials \textit{via} the self-assembly of building blocks on the nanoscale presents an enormous opportunity for the design of materials with targeted behavior, including mechanical and optical properties.\cite{Jain2014a,Boles2016a}
Following the definition of Whitesides \textit{et al.},\cite{Whitesides2002} a self-assembly process is characterized by the emergence of structure from disordered, distinct constituents and governed by their shapes and interactions.
In order to design a material for synthesis \textit{via} self-assembly, we need to answer the question ``What constituents are required for the targeted self-assembly behavior?''
This question represents the \emph{inverse} problem in contrast to the \emph{forward} problem of ``What is the self-assembly behavior of certain predefined constituents?''~\cite{Glotzer2004,Jain2014a}
The major challenge in solving the \emph{inverse} problem is the vast search space constituted by the sheer limitless choice and possible combinations of feasible building blocks and interactions.\cite{Glotzer2007,Torquato2009,Cademartiri2012}
Of course, simply identifying the constituents that produce a thermodynamic target structure does not guarantee the existence of a robust kinetic pathway to that structure. 

Although directing self-assembly processes with highly specific interactions is technically possible,\cite{Jacobs2015d,Jacobs2016,Huntley2016} it is often more informative to know what is the \emph{simplest} interaction needed to achieve a specific structure \textit{via} facile and robust self-assembly,\cite{Kumar2017} that is, on short time scales and without the need for seeding the target crystal. This so-called simplest interaction will not only provide insight into the underlying mechanisms of self-assembly, but may also be easier to realize experimentally and produce higher yields.
Simple interactions with features whose length scales are on the order of the interparticle distances are experimentally realizable through, for example, DNA-mediated surface functionalization of nanoparticles.\cite{Knorowski2011,Li2012,Macfarlane2013,Li2013b,Auyeung2014,OBrien2016a,Lin2017a,Wang2017f}

In this work we optimize \glspl{ipp} as a model for the interaction between point particles that self-assemble into a specific target crystal structure from a fluid (disordered) state.
That is, we seek pair potentials that not only have shapes containing minimal features, but also which drive assembly of the target structure rapidly, without need for a seed and without long waiting times for nucleation.
It was previously shown that Fourier space filters provide an elegant way to \emph{design} simple \glspl{ipp} for the self-assembly of complex structures.\cite{Edlund2011,Edlund2013a}
Here we apply this knowledge to advance the \gls{remin} approach outlined by Lindquist \textit{et~al.~}\cite{Lindquist2016b} to be carried out directly in Fourier space and with the repeated application of a smooth low-pass filter at each iteration in order to effectively steer the optimization process towards simpler solutions.
The proposed \gls{ffrem} method (Fig.~\ref{fig:method}) is designed to optimize for potentials without the need to restrict the range of interactions fed into the algorithm or limit the solution to a specific parametrization.
Instead, a low-pass filter is imposed in reciprocal space that penalizes features in the potential at large $k$. This filtering leads to effectively fewer minima and maxima and the suppression of noisy fluctuations on length scales smaller than those features in real space while naturally preserving the real-space potential range and qualitative functional form.

It is known that the robustness of convergence of standard methods for the derivation of \glspl{ipp} for \emph{fluids},\cite{Soper1996,Reith2003,Ruhle2009,Sanyal2016,Ingolfsson2013} many of which fall under the general umbrella of the relative entropy minimization framework,\cite{Shell2008,Chaimovich2011} can be improved through smoothing directly in real-space.
However, \glspl{rdf} of solids have many more characteristic length scales compared to their fluid counterparts making it especially difficult to converge non-parametrized solutions that are neither over- nor underfitted without a judicious choice of cut-off and smoothing filter.
That means in this context that they contain too many features and length scales that are not actually critical and are possibly even detrimental for the robust self-assembly of the target structure. 
Since we know that complex structures may be assembled from much simpler potential functions,\cite{Engel2014} an efficient optimization algorithm needs to be biased towards those length scales that are essential for robust self-assembly.

Another approach to steer the optimization of potentials towards simpler solutions is to apply constraints, \textit{e.g.}, by limiting the solution space to a specific functional form.\cite{Lindquist2016}
Overfitting may also be prevented with early stopping for more broadly constrained search spaces, for example when the solutions are limited to a specific class of functions, such as repulsive, monotonically decreasing functions~\cite{Lindquist2016b,Jadrich2017} or parameterized splines that effectively implement a lower limit on all feature length scales.\cite{Lindquist2018}
The \gls{ffrem} method does not rely on such constraints, but instead steers the optimization towards smoother and simpler solutions by the repeated application of a filter function in Fourier space ($k$-space).
This approach is especially advantageous during early exploration, \textit{e.g.}, to determine whether any solution exists at all, or when there is no specific desired functional form.
Conversely, the presented method does not allow one to target a specific functional form, even if desired.

\section{Fourier-filtered Relative Entropy Minimization}

For the algorithm's derivation we recognize the potential energy $E$ of a three-dimensional system of interacting point particles in a volume $V=N/\rho$, where $\rho$ denotes the number density, may be expressed as a function of the \gls{rdf}, $g(r)$, both in real space,

\begin{align}
    \frac{E}{N} &= 2 \pi \rho \int_{0}^\infty \diff{r} \, r^2 g(r) V(r) \text{,}
\end{align}
and equivalently in reciprocal space,

\begin{align}
\frac{E}{N} &= 2 \pi \rho \int_{0}^\infty \diff{k} \, k^2 \hat g(k) \hat V(k) \text{,}
\end{align}
where $f(r) \mapsto \hat f(k)$ is the Fourier transform, defined by

\begin{align}
    \hat f(k) = \frac{1}{k}\sqrt{\frac{2}{\pi}} \int_0^\infty \diff{r} \, r f(r) \sin \left(kr\right) \text{.}
\end{align}
and $V(r)$ and $\hat V(k)$ represent the isotropic pairwise interaction potential in real- and $k$-space respectively.

The Fourier transformation is unique and invertible and thus preserves all the information of the real-space potential.
However, in practice, in order to meet the complexity constraints introduced above, a real-space potential is strongly limited, especially in its range.
This means that traditional optimization techniques---carried out exclusively in real-space---are inherently tying the information exploited for the optimization process to the range of the potential energy function.
In other words, a potential optimized with, \textit{e.g.}, \gls{ibi} is inherently biased to match short-range distance distributions since any long-range information contained in the \gls{rdf} beyond the real-space potential cut-off is completely discarded.
By instead optimizing the pairwise interaction model directly in Fourier space, we introduce no inherent constraint on the potential range and the potential function is only transformed into real space for the sake of carrying out the integration of forces as part of simulating the assembly process using \gls{md}.

For the overall process (shown in Fig.~\ref{fig:method}), we first propose an \textit{ansatz} function $\hat V^{(0)}(k)$, which in our case is just the smoothened Fourier transform of the potential of mean force.
Then we enter an iterative update process, where at each iteration we map the potential to real space and carry out a \gls{md} simulation of point particles.
Specifically, we thermalize the system at an elevated temperature of $k_B T=3.0\varepsilon$ to ensure that it is in a disordered fluid state, and then cool and compress the system over the next 4 million time steps to a final temperature of $k_B T=1.0 \varepsilon$.
Whether the system assembled the targeted structure or not, we then calculate the shifted \gls{rdf} $h(r) = g(r) - 1$ and Fourier transform to obtain $\hat h(k)$.
The update step is then derived from the minimization of relative entropy directly in Fourier space and is expressed as a function of the difference between $\hat h^{(i)}(k)$ at iteration $i$ and $\hat h^*(k)$ measured from the target structure

\begin{align}
    \hat V^{(i+1)}(k) &= \hat V^{(i)}(k) + \alpha \me^{-c k} k_B T [\hat h^{(i)}(k) - \hat h^*(k)] \label{eq:update-step} \text{,}
\end{align}
where $\alpha$ denotes the effective learning rate, $c$ scales the low-pass filter, and $k_B T$ is the thermal energy of the system.
The learning rate $\alpha$ is a unitless dampening factor to stabilize the optimization process; we found values on the order of 0.1 to be small enough to yield stable optimization.
The low-pass length scale of the exponential filter is set by $c$ such that features in the real-space potential with wavelengths much smaller than $2\pi c$ are damped while features with much longer wavelengths are preserved.
The studied filter strengths $c=0.1\sigma$ and $c=0.2\sigma$ are chosen empirically, such that features on length scales on the order of particle interactions $\mathcal{O}(1)$ are largely preserved, while features on smaller length scales are sufficiently suppressed.

\begin{figure*}
    \centering
    \includegraphics[width=0.90\linewidth]{{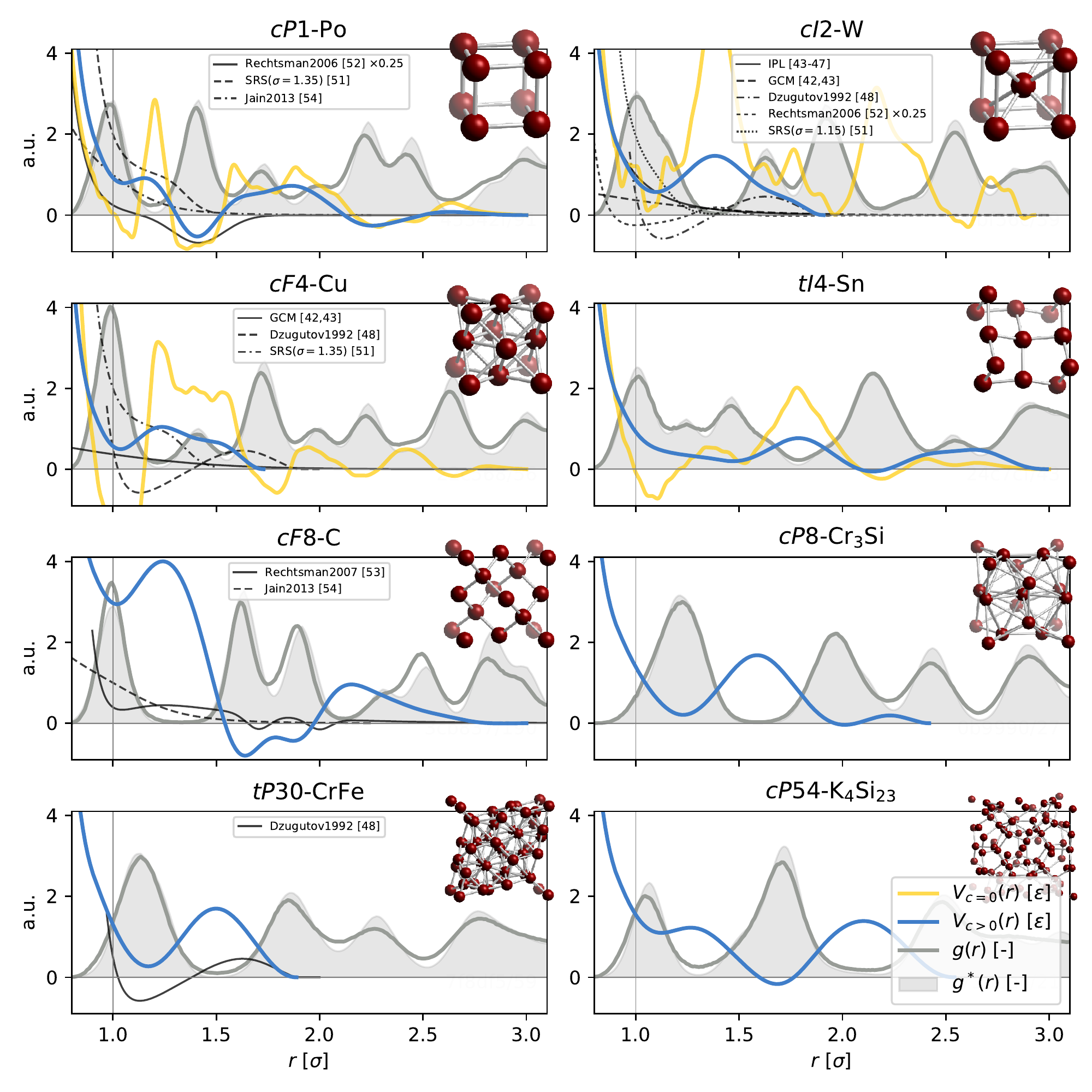}}
    \caption{
        Here we show the objectively best \glspl{ipp} (corresponding to $\objective_\text{max}$, see Eq.~\ref{eq:objective}) optimized with \gls{ffrem} ($c>0$) in blue and without filtering ($c=0$) in yellow.
        The \gls{rdf} measured from the assembled structures ($g(r)$, gray lines) are compared against those obtained from the target harmonic crystal ($g^*(r)$, shaded in gray).
        We found that a perfect fitting of all \gls{rdf} features is not a critical requirement for the assembly of the target structure. 
        The corresponding unit cells are depicted as ball-and-stick models~(top right).
        While our method consistently produces simpler potentials compared to the control method without filtering, it is not guaranteed that our methodology results in the simplest possible interaction potential for a given target structure.
        This becomes obvious in comparison with select results from the literature, where some potentials are significantly simpler compared to our results, even though most share general characteristics.
        The potentials drawn from the literature  and plotted here are not adjusted for differences in temperature and density of the assembly state point.
    }
    \label{fig:overview}
\end{figure*}

\Glspl{ipp} mapped from Fourier space onto real space need to be truncated since they are in principle infinite in range.
For this we applied the following cut-off algorithm:

\begin{align}
    r_\text{cut} &= \min_r \left(
        r \geq r_\mathrm{min}
        \land V(r) \leq \epsilon
        \land V'(r) \leq \epsilon' \right)
    \label{eq:truncation}
    \text{,}
\end{align}
where we chose $r_\text{min} \in \{1.6, 2.4\}$, $\epsilon=0.3$ and $\epsilon'=5.0$.
This means that the potential is cut off at the first extremum beyond $r_\mathrm{min}$ that is sufficiently close to zero.
We ensure smoothness at this cut-off by applying the Stoddard-Ford algorithm~\cite{Stoddard1973} up to the first derivative.

To apply this algorithm to the derivation of \glspl{ipp} for the assembly of solid structures, we compute the \gls{rdf} from position distributions of harmonic crystals, where particles are bound to their ideal crystal sites through harmonic bonds similarly to methods previously described in the literature.\cite{Lindquist2016b,Lindquist2018}
The harmonic bond constant $K$ was chosen such that the peaks within the measured \gls{rdf} are sufficiently distinct to reliably characterize the structure, usually in a range of $K=[100,800]$, but always low enough to avoid singularities.

All molecular dynamics simulations were carried out with HOOMD-blue~\cite{Anderson2008,Glaser2014b} on XSEDE resources~\cite{Towns2014} (including the Comet and Bridges clusters) and on the high-performance compute cluster of the University of Michigan.
The computational workflow in general and data management in particular for this publication was primarily supported by the signac data management framework.\cite{Adorf2018}

Simulation trajectories were analyzed with the software package freud~\cite{freud} and visualizations were rendered with CrystalMaker and Fresnel.\cite{fresnel}
Structures were analyzed and identified with the in house software Injavis.
We trained a machine-learning model based on a deep neural network with spherical harmonic descriptors of particle environments to identify crystal structures from millions of simulation snapshots.\cite{Spellings2018}

To benchmark the performance of \gls{ffrem}, we also attempted a control optimization using standard \gls{remin}, which is equivalent to no filtering ($c=0$).
\Glspl{ipp} optimized using \gls{remin} without any kind of filtering failed to self-assemble the target structure in about 70\% of all cases.

\section{Isotropic Pair Potentials for Complex Structures}

Design and optimization of \glspl{ipp} for simple and complex structures has yielded a plethora of different models ranging from repulsive to attractive, from short-ranged to long-ranged, from simple to complex.
We have selected a few exemplary models to compare our results to, including the \gls{gcm},\cite{Stillinger1976,Prestipino2005b} the \gls{ipl},\cite{Domb1951,Hoover1971,Laird1992,Agrawal1995,Prestipino2005b} the Dzugutov potential,\cite{Dzugutov1992,Dzugutov1993,Roth2000} the \gls{srs},\cite{Fomin2008} and potentials published by Rechtsman \textit{et al.},\cite{Rechtsman2006a,Rechtsman2007a} Jain \textit{et al.},\cite{Jain2013} and recently by Lindquist \textit{et al.}\cite{Lindquist2018} that have been shown to assemble some of the structures we targeted as part of this study, and many of which share qualitative characteristics with our results.

Using \gls{ffrem} we found \glspl{ipp} for the assembly of \gls{cP1}, \gls{cI2}, \gls{cF4}, \gls{betaSn}, \gls{cP8}, \gls{cF8}, \gls{cP54}, and \gls{sigmaphase} structures.
The corresponding \glspl{ipp} are plotted in Fig.~\ref{fig:overview}.
Without filtering, \textit{i.e.}, $c=0$, we were only able to optimize potentials for \gls{cP1}, \gls{cI2}, \gls{cF4}, and \gls{betaSn}.
Notably, the potentials we found for the Frank-Kasper phases (\gls{cP8}, \gls{sigmaphase}) are highly similar to those reported by Lindquist and co-workers\cite{Lindquist2018} obtained with standard \gls{remin} in combination with a spline interpolation.

Attempts to find potentials for \gls{cP4}, \gls{betaMn}, and \gls{cI52} were not successful with parameters tested for this study, that means they did not assemble the target structure after a fixed number of time steps.
This does not rule out the possibility of the obtained potentials to self-assemble the target structure using alternative protocols, \textit{e.g.}, by starting from a seeded configuration or simply sampling longer to overcome potential nucleation barriers.
This is evidenced by the fact that potentials that will self-assemble the targeted structures are known for all tested structures, including \gls{cI52}\cite{Zetterling2000} and \gls{betaMn},\cite{Elenius2009} and because the assembly yield is equal or greater for all potentials when the system is doped with a crystalline seed.
Within the realm of this study, we only report those potentials that assemble the target structure with the tested protocol, others were considered unsuccessful and consequently disregarded.

To quantify the effectiveness of our filtering, we introduced a measurement of complexity, $\Omega$, defined as

\begin{align}
    \Omega &\equiv \frac{1}{k_\text{max}} \int_{k=0}^{k_\text{max}} \diff{k} \, \left[ k \hat V(k) \right]^2 \label{eq:complexity}\text{.}
\end{align}
$\Omega$ is nonnegative and becomes large when the potential has small-scale real-space features.

\begingroup
\begin{table}
\caption{The complexity $\Omega$ as defined by Eq.~\ref{eq:complexity}, measured for different structures and filter strengths $c$ (Eq.~\ref{eq:update-step}).}
\begin{ruledtabular}
\begin{tabular}{lrrrr}
Crystal Structure &   $c=0$ &   $c=0.1\sigma$ &   $c=0.2\sigma$ &  \textbf{Mean} \\
\hline
$cP1$-$\chem{Po}$              &  1.04 &  0.22 &  0.18 &  0.48 \\
$cI2$-$\chem{W}$               &  0.46 &  0.08 &  0.06 &  0.20 \\
$cF4$-$\chem{Cu}$              &  0.99 &  0.06 &  0.04 &  0.37 \\
$tI4$-$\chem{Sn}$              &  0.71 &  0.07 &  0.04 &  0.27 \\
$cF8$-$\chem{C}$               &   - &   - &  0.70 &  0.70 \\
$cP8$-$\chem{Cr_3Si}$           &   - &   - &  0.09 &  0.09 \\
$tP30$-$\chem{CrFe}$           &   - &   - &  0.06 &  0.06 \\
$cP54$-$\chem{K_4Si_{23}}$         &   - &   - &  0.07 &  0.07 \\
\end{tabular}
\end{ruledtabular}
\label{tab:complexity}
\end{table}
\endgroup

The effectiveness of the low-pass filter becomes obvious when comparing $\Omega$ between optimization procedures with different filter strengths $c$, see Tab.~\ref{tab:complexity}.
Optimization runs with the higher $c$ value of $0.2\sigma$ consistently yielded solutions with lower complexity.

\begin{figure}
    \centering
    \includegraphics{{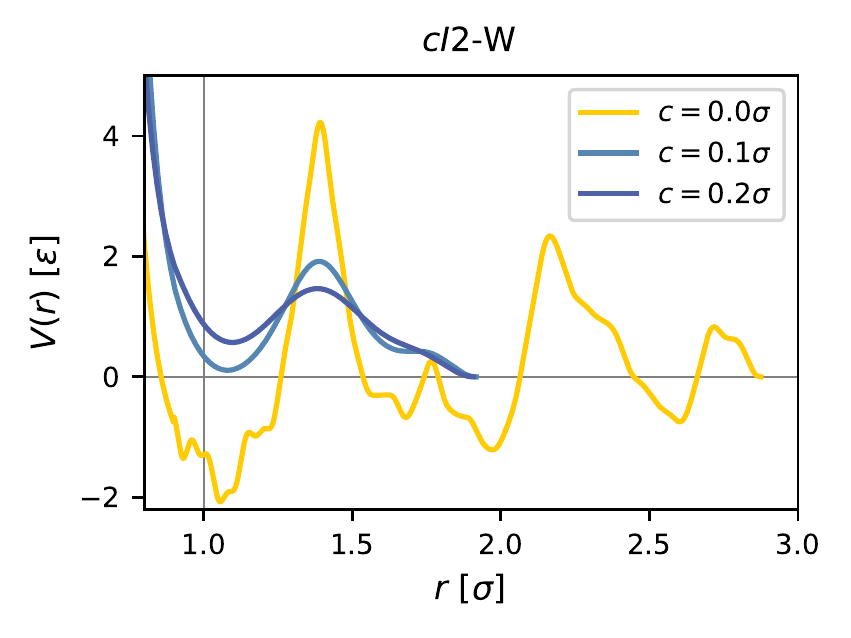}}
    \caption{
        We can use the filter strength $c$ to effectively control the complexity of our solution in $k$-space (Eq.~\ref{eq:update-step}), resulting in smoother potentials with fewer features on smaller length scales.
        Important features such as the location of extrema and their relative well-depth are preserved.
        The control optimization with $c=0$ is obviously much more complex.
    }
    \label{fig:compare-opt-parameters}
\end{figure}

This effect can be visualized when comparing solutions mapped onto real~space for the identical target structure, but carried out with different filter strengths.
Fig.~\ref{fig:compare-opt-parameters} shows solutions for \gls{cI2}, optimized with $c$ ranging from $0$ to $0.2\sigma$.
The solution for $c=0$ is clearly much more complex compared to all other solutions, but the main characteristics of the potential functions with $c>0$ are conserved.
The solution with $c=0.1\sigma$ clearly contains additional non-critical features on smaller length scales compared to solutions obtained with $c=0.2\sigma$.

\begin{figure}
    \centering
    \includegraphics{{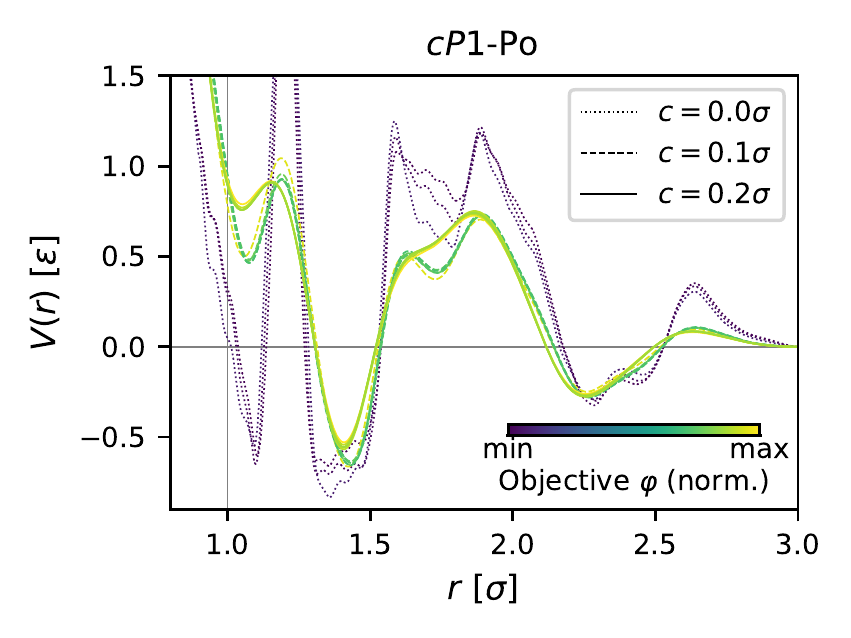}}
    \caption{
        All optimized potentials were evaluated by the objective function $\objective$ (defined by Eq.~\ref{eq:objective} and visualized by color), which rewards a better  fit with the target \gls{rdf} and penalizes complexity (Eq.~\ref{eq:complexity}) and long-ranged potentials (Eq.~\ref{eq:truncation}).
        Plotted are potentials that were found as solutions for the self-assembly of the \gls{cP1} structure for a given set of optimization parameters and multiple replications.
        The potentials have about the same shape, but the complexity is sharply clustered with respect to the filter strength $c$.
    }
    \label{fig:objective-function}
\end{figure}

\section{Method Evaluation}

To quantify the objective of minimizing the difference in the \gls{rdf} measured from the thermalized ideal crystal and the distribution measured from the self-assembly result, with the additional constraint of minimizing complexity and potential range, we define the objective function,

\begin{align}
    \objective^{(i)} &= \frac{f^{(i)}}{\Omega^{(i)} \cdot r^{(i)}_\text{cut}} \text{,} \label{eq:objective}
\end{align}
where

\begin{align}
    f^{(i)} &= 1.0 - \frac{\sum_j |g^{(i)}(r_j)-g^*(r_j)|}{\sum_j [g^{(i)}(r_j) + g^*(r_j)]} \label{eq:fitness} \text{.}
\end{align}
The functions $g^{(i)}(r_j)$ and $g^*(r_j)$ denote the discrete \gls{rdf} measured at iteration $i$ and from the target structure, respectively; therefore the fitness $f$ is a measure of how closely the \gls{rdf} matches the target distribution at iteration~$i$.
The measure $f$ is normalized by the magnitude of the compared values so that different systems are comparable.\cite{Moore2014}
This means that the objective function naturally increases as the \gls{rdf} matches better, but is reduced by increased complexity (Eq.~\ref{eq:complexity}) and the range of the potential evaluated in real~space (Eq.~\ref{eq:truncation}).
The objective function is used to rank different \gls{ipp} solutions for the same structure as shown in Fig.~\ref{fig:objective-function}.

For the structures that we were able to find a solution for, the Fourier space filtering not only results in a reduction of \gls{ipp} complexity, but also generally yields an overall higher fitness.
While the literature suggests that an increased filter strength would result in improved robustness of the optimization process and the assembly kinetics,\cite{Reith2003,Ruhle2009} we found it both surprising and reassuring that the least complex solutions result in an overall better fitting of the target function as well.

In general, the fitness of the resulting \glspl{rdf} is not a good indicator for successful assembly.
In fact, when evaluated for the complete data set, the fitness is only weakly correlated with the yield.
In other words, most optimization runs returned potentials that reliably reproduced the \gls{rdf}, up to a specific precision, but from that alone we cannot discern that this potential assembles the target structure, or any ordered structure at all.
However, for those potentials that did assemble the target structure, we can use the fitness as a quantitative measure of how well the solution matches the target distribution, which in turn allows us to rank multiple successful solutions.

While we were able to determine \glspl{ipp} for many different structures, in many cases the optimization resulted in potentials that either failed to assemble any structure, \textit{i.e.}, they formed some kind of fluid, or assembled highly defective structures that might or might not resemble the target structure.
In only very few cases did the optimization result in a structure different from the desired structure.
For example, none of the attempts to optimize an \gls{ipp} for \gls{cI52} were successful, but some of them yielded \gls{cI2} instead.
While the optimization did not succeed in this case, it at least yielded a closely related state: \gls{cI52} represents a $(3\times3\times3)$-fold superstructure of \gls{cI2}.
Similarly, some of the \gls{betaMn} optimization runs yielded \gls{cI2} as well.
A potential energy analysis shows that the \gls{ffrem} algorithm is able to determine a potential for which the targeted structure presents the ground state among the competitor pool in all cases except for \gls{cI52}, where \gls{cI2} has a lower potential energy.
We therefore presume that the failure to find a \gls{ipp} for a targeted structure is related primarily to the assembly protocol in all but this case.

We analyzed the optimization performance with respect to the optimization parameters and with respect to the target structures.
In particular we are interested in determining which parameters yield the best results and whether we can discern for which structures it is inherently harder to optimize potentials with the presented methods based on specific structural characteristics.

We evaluated the robustness of a specific parameter and structure combination by dividing the number of times they resulted in the successful optimization of a potential in at least one iteration with the total number of attempts for that combination.
All combinations were replicated independently three times with a different random seed.

\begin{figure}
    \centering
    \subfloat[Optimization yield by structure]{
        \includegraphics{{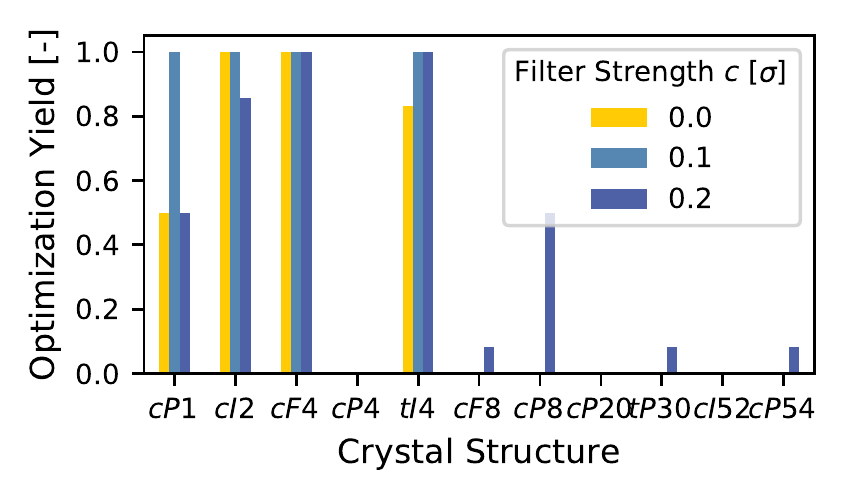}}
        \label{fig:yield-optimization-by-structure}
        }\\
    \subfloat[Optimization yield by filter strength $c$]{
        \centering
        \includegraphics{{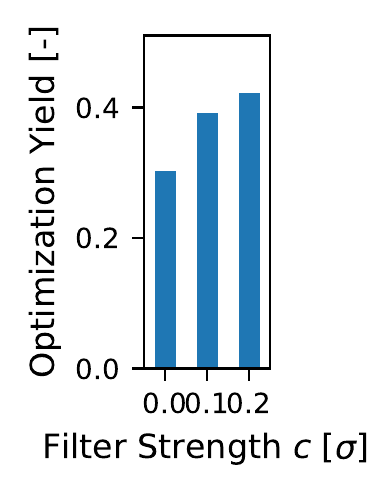}}
        \label{fig:yield-optimization-by-c}
        }
    \subfloat[Optimization yield by\newline unit cell size]{
        \centering
        \includegraphics{{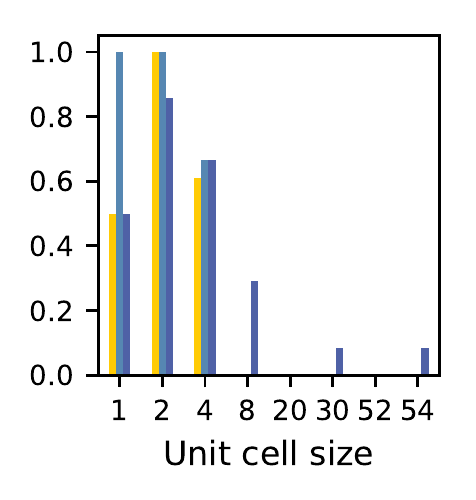}}
        \label{fig:yield-optimization-by-unit-cell-size}
        }
    \caption{
        The yield shown here is the number of successful attempts, \textit{i.e.}, a structure and optimization parameter combination, where at least one iteration led to an \gls{ipp} that would assemble the target structure, divided by the total number of attempts.
    }
    \label{fig:yield-optimization}
\end{figure}

We found that the overall yield of \gls{ffrem} (41\%) for finite filter strengths ($c>0$) is higher compared to 30\% with no filtering ($c=0$).
The yield among replication groups, that means all attempts with identical optimization parameters except for the random seed,  where at least one attempt led to successful optimization, is much closer (87\% (filtered) versus 95\% (non-filtered)).

For the three tested filter strengths $c$, we find that \gls{ffrem} performs better with higher values of $c$ for almost all structures except for \gls{cP1} and \gls{cI2} (Fig.~\ref{fig:yield-optimization-by-structure}) as well as the overall average (Fig.~\ref{fig:yield-optimization-by-c}).
This finding is surprising to us as we expected that increased smoothing---that means seemingly less preserved information---might make it more difficult to assemble more complex target structures with larger unit cells.
Overall we were able to find more structures more robustly with \gls{ffrem} in the vast majority of studied cases.

It appears that for the tested structures the yield is negatively correlated with the unit cell size (see Fig.~\ref{fig:yield-optimization-by-unit-cell-size}).
However, we would need to test more structures to determine whether this is inherent to the optimization algorithm, or whether, \textit{e.g.}, it is because structures with larger unit cells are generally more difficult to assemble with the used protocols and without seeds.
Furthermore, the yield increases with the average coordination number of the first neighbor shell.
With the exception of the extraordinarily robust optimization of potentials for the \gls{cP1} and \gls{betaSn} structures, which both have an average coordination number of 6, there appears to be a a positive relationship between average coordination number and the yield.
Specifically, it appears to be generally more difficult to optimize \glspl{ipp} for lower-coordinated structures, such as \gls{cP4} or \gls{cF8} (see Fig.~\ref{fig:yield-optimization-by-structure}).

\begin{figure}
    \centering
    \includegraphics{{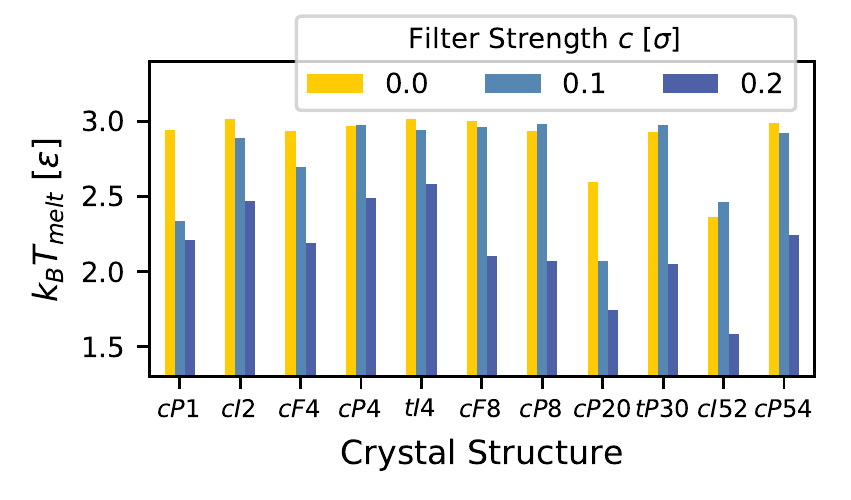}}
    \caption{
        The melting temperature $k_{B}T_\text{melt}$ is the temperature at which the target structure disintegrates, \textit{i.e.}, is no longer stable with particles at the lattice coordinates interacting \textit{via} the optimized \gls{ipp}.
        The melting temperature was determined by initializing the system in the target structural configuration and then slowly raising the temperature from $k_{B}T=1.0$ up to $k_{B}T=3.0$.
    }
    \label{fig:melting-temperature}
\end{figure}

To assess the optimization robustness for all runs, even for those parameter combinations where the target structure could not be assembled, we determined the melting temperature $k_{B}T_\mathrm{melt}$ for the best potential, \textit{i.e.}, the potential with the objectively highest value ($\objective_{\max}$) (see Fig.~\ref{fig:melting-temperature}).
The melting temperature is defined as the temperature at which the target structure comprised of particles interacting \textit{via} the corresponding potential would start to disintegrate.
We initialized simulation configurations with the target structure, and then slowly increased the temperature while evaluating the Lindemann criterion:\cite{lindemann1910,Gilvarry1956}

\begin{align}
    L &= {\left(\mean{\left(\vec{r}_i - \vec{r}_0\right)}\right)}^{1/2}
    \text{.}
\end{align}
The melting temperature was then determined to be exactly the temperature at which $L$ and $\frac{\delta L}{\delta t}$ were above a specific but universal threshold.

We found that the mean melting temperature is slightly higher with lower filter strengths $c$.
The mean melting temperature for individual structures was often comparable or even higher with no filter ($c=0$), even when the latter did not result in a potential that would assemble the target structure.

In conclusion, we demonstrated that by taking advantage of the unique properties of Fourier space, we are able to implement a simple but effective optimization algorithm resulting in smooth \glspl{ipp} for the self-assembly of complex structures.
\Gls{ffrem} is more robust and we demonstrate an overall higher yield compared to the control optimization without a filter.
In addition, the smoother potentials resulting from \gls{ffrem} lead to a significantly higher yield in the self-assembly simulations of complex crystal structures.

\section*{Source Code}

The source code for the execution of \gls{ffrem} optimizations with the simulation package HOOMD-blue can be downloaded at \url{http://glotzerlab.engin.umich.edu/ff-rem}.

\begin{acknowledgments}
We thank Paul M.\ Dodd, Pablo F.\ Damasceno, and Bryan J.\ VanSaders for helpful discussions and Rose K.\ Cersonsky for assistance with structure identification.
We thank Matthew Spellings for support with the implementation of the machine-learning based crystal structure detection algorithm.
This research was supported in part by the National Science Foundation, Division of Materials Research Award \# DMR 1409620 and by a Simons Investigator Award from the Simons Foundation to Sharon Glotzer. J.A.\ is supported by a National Science Foundation Graduate Research Fellowship Grant No.\ DGE 1256260.
J.D.\ acknowledges support through the Early Postdoc.Mobility Fellowship from the Swiss National Science Foundation, grant number P2EZP2\_152128.
This work used the Extreme Science and Engineering Discovery Environment (XSEDE), which is supported by National Science Foundation grant number ACI-1053575; XSEDE award DMR 140129.
Additional computational resources and services provided by Advanced Research Computing at the University of Michigan, Ann Arbor.
\end{acknowledgments}

\bibliography{ipp-k-space-optimization-paper}
\end{document}